\documentclass[preprint,3p,sort&compress]{elsarticle}

\usepackage{amsmath,amssymb}

\newcommand{\Eps}{\varepsilon}
\newcommand{\Dmq}{\Delta m^2}
\newcommand{\diag}{\mathop{\mathrm{diag}}}
\renewcommand{\Re}{\mathop{\mathrm{Re}}}
\renewcommand{\Im}{\mathop{\mathrm{Im}}}

\begin{document}

\begin{flushright}
  IFT-UAM/CSIC-16-048\\
  YITP-SB-16-23
\end{flushright}

\title{Non-standard Neutrino Interactions in the Earth and the Flavor
  of Astrophysical Neutrinos}

\author[sb,ub]{M.~C.~Gonzalez-Garcia}
\ead{maria.gonzalez-garcia@stonybrook.edu}

\author[uam]{Michele Maltoni}
\ead{michele.maltoni@csic.es}

\author[uam]{Ivan Martinez-Soler}
\ead{ivanj.m@csic.es}

\author[sb]{Ningqiang Song}
\ead{ningqiang.song@stonybrook.edu}

\address[sb]{C.N.~Yang Institute for Theoretical Physics, SUNY at
  Stony Brook, Stony Brook, NY 11794-3840, USA}

\address[ub]{Instituci\'o Catalana de Recerca i Estudis Avan\c{c}ats
  (ICREA), Departament d'Estructura i Constituents de la Mat\`eria and
  ICC-UB, Universitat de Barcelona, 647 Diagonal, E-08028 Barcelona,
  Spain}

\address[uam]{Instituto de F\'{\i}sica Te\'orica UAM/CSIC, Calle de
  Nicol\'as Cabrera 13--15, Universidad Aut\'onoma de Madrid,
  Cantoblanco, E-28049 Madrid, Spain}

\begin{abstract}
  We study the modification of the detected flavor content of ultra
  high-energy astrophysical neutrinos in the presence of non-standard
  interactions of neutrinos with the Earth matter. Unlike the case of
  new physics affecting the propagation from the source to the Earth,
  non-standard Earth matter effects induce a dependence of the flavor
  content on the arrival direction of the neutrino.  We find that,
  within the current limits on non-standard neutrino interaction
  parameters, large deviations from the standard $3\nu$ oscillation
  predictions can be expected, in particular for fluxes dominated by
  one flavor at the source. Conversely they do not give sizable
  corrections to the expectation of equalized flavors in the Earth
  for sources dominated by production via pion-muon decay-chain.
\end{abstract}

\begin{keyword}
  Astrophysical neutrinos, non-standard neutrino interactions.
\end{keyword}

\maketitle

\section{Introduction}

The detection of ultra-high energy neutrinos of astrophysical origin
in IceCube~\cite{Aartsen:2013bka, Aartsen:2013jdh, Aartsen:2014gkd,
  Aartsen:2015rwa} marks the begin of high energy neutrino astronomy.
From the point of view of astronomy, the main open question resides in
finding the sources of such neutrinos, an issue to which many
suggestions have been contributed (for a recent review see
Ref.~\cite{Anchordoqui:2013dnh}).  More on the astrophysical front,
one also questions what type of mechanisms are at work in those
sources to produce such high energy neutrino flux.  To address this
question the measurement of the flavor composition of the observed
neutrinos acquires a special relevance.  For example, for the
pion-muon decay chain, which is the most frequently considered, one
expects $\phi^s_\mu = 2\phi^s_e$ while $\phi^s_\tau =
0$~\cite{Learned:1994wg} (denoting by $\phi^s_\alpha$ the neutrino
flux of flavor $\nu_\alpha$ at source). Alternatively, if some of the
muons lose energy very rapidly one would predict a single $\mu$-flavor
flux while $\phi^s_e = \phi^s_\tau = 0$~\cite{Kashti:2005qa,
  Lipari:2007su, Kachelriess:2007tr, Hummer:2010ai, Winter:2014pya}.
If neutrino production is dominated by neutron decay one expects also
a single flavor flux but of electron neutrinos~\cite{Lipari:2007su} so
in this case $\phi^s_\mu = \phi^s_\tau = 0$. Decay of charm mesons
contribute a flux with equal amounts of electron and muon neutrinos,
$\phi^s_e = \phi^s_\mu$ and $\phi^s_\tau = 0$.  If several of the
above processes in the source compete, arbitrary flavor compositions
of $\phi^s_e$ and $\phi^s_\mu$ are possible but still with $\phi_\tau
= 0$~\cite{Hummer:2010ai}. If, in addition, $\nu_\tau$ are also
produced in the source~\cite{Lunardini:2000fy, Razzaque:2009kq,
  Sahu:2010ap}, then generically $\phi^s_\alpha \neq 0$ for $\alpha =
e, \mu, \tau$.

Neutrino oscillations modify the flavor composition of the neutrino
flux by the time they reach the Earth. In the context of the well
established framework of $3\nu$ oscillations these modifications are
well understood and quantifiable given the present determination of
the neutrino oscillation parameters.  Because of this several studies
to quantify the flavor composition of the IceCube events, even with
the limited statistics data available, have been
presented~\cite{Mena:2014sja, Bustamante:2015waa, Palladino:2015zua,
  Palomares-Ruiz:2015mka, Watanabe:2014qua, Kawanaka:2015qza,
  Aartsen:2015knd, Vincent:2016nut} but the results are still
inconclusive.

It is well-known that new physics (NP) effects beyond $3\nu$
oscillations in the neutrino propagation can alter the predicted
flavor composition of the flux reaching the Earth, thus making the
task of elucidating the production mechanism even more
challenging. Examples of NP considered in the literature include
Lorentz or CPT violation~\cite{Hooper:2005jp}, neutrino
decay~\cite{Beacom:2002vi, Baerwald:2012kc}, quantum
decoherence~\cite{Anchordoqui:2005gj, Hooper:2004xr} pseudo-Dirac
neutrinos~\cite{Beacom:2003eu, Esmaili:2009fk}, sterile
neutrinos~\cite{Athar:2000yw}, non-standard neutrino interactions with
dark matter~\cite{deSalas:2016svi}, or generic forms of NP in the
propagation from the source to the Earth parametrized by effective
operators~\cite{Arguelles:2015dca}. Besides modifications of the
flavor ratios many of these NP effects also induce a modification of
the energy spectrum of the arriving neutrinos.

In this paper we consider an alternative form of NP, namely the
possibility of non-standard interactions (NSI) of the neutrinos in the
Earth matter. Unlike the kind of NP listed above, this implies that
neutrinos reach the Earth surface in the expected flavor combinations
provided by the ``standard'' $3\nu$ vacuum oscillation mechanism: in
other words, NSI in the Earth affect only the flavor evolution of the
neutrino ensemble from the entry point in the Earth matter to the
detector. The goal of this paper is to quantify the modification of
the neutrino flavor composition at the detector because of this effect
within the presently allowed values of the NSI parameters. To this aim
we briefly review in Sec.~\ref{sec:forma} the formalism employed and
derive the relevant flavor transition probabilities from the source to
the detector including the effect of NSI in the Earth.  We show that
the resulting probabilities are energy independent while they depend
on the zenith angle arrival direction of the neutrinos, in contrast
with NP affecting propagation from the source to the Earth.  Our
quantitative results are presented in Sec.~\ref{sec:results}, where in
particular we highlight for which source flavor composition the
Earth-matter NSI can be most relevant. Finally in
Sec.~\ref{sec:conclu} we draw our conclusions.

\section{Formalism}
\label{sec:forma}

Our starting point is the initial neutrino (antineutrino) fluxes at
the production point in the source which we denote as $\phi^s_\alpha$
($\bar\phi^s_\alpha$) for $\alpha = e, \nu, \tau$. The corresponding
fluxes of a given flavor at the Earth's surface are denoted as
$\phi^\oplus_\alpha$ ($\bar\phi^\oplus_\alpha$) while the fluxes
arriving at the detector after traversing the Earth are
$\phi^d_\alpha$ ($\bar\phi^d_\alpha$). They are generically given by
\begin{equation}
  \label{eq:eq1}
  \phi^\oplus_\beta(E)
  = \sum_\alpha \int dE' \mathcal{P}^{s\to \oplus}_{\alpha\beta}(E,E')
  \phi^s_\alpha(E') \,,
  \qquad
  \phi^d_\beta(E)
  = \sum_\alpha \int dE'\mathcal{P}^{s\to d}_{\alpha\beta}(E,E')
  \phi^s_\alpha(E')
\end{equation}
and correspondingly for antineutrinos. $\mathcal{P}$ is the flavor
transition probability including both coherent and incoherent effects
in the neutrino propagation.

\subsection{Coherent effects}

Let us start by considering first only the coherent evolution of the
neutrino ensemble. In this case, the flavor transition probabilities
from the source ($s$) to the Earth entry point ($\oplus$) and to the
detector ($d$) can be written as
\begin{align}
  \label{eq:pse}
  \mathcal{P}^{s\to \oplus}_{\alpha\beta}(E,E')
  &= P^{s\to \oplus}_{\alpha\beta}(E) \, \delta(E-E') \,,
  & \text{with}\quad
  P^{s\to \oplus}_{\alpha\beta}(E)
  &= \left|A^{s\to \oplus}_{\alpha\beta}(E)\right|^2
  \\
  \mathcal{P}^{s\to d}_{\alpha\beta}(E,E')
  &= P^{s\to d}_{\alpha\beta}(E) \, \delta(E-E') \,,
  & \text{with}\quad
  P^{s\to d}_{\alpha\beta}(E)
  &= \left|A^{s\to d}_{\alpha\beta}(E)\right|^2
  = \left| \sum_\gamma A^{s \to \oplus}_{\alpha\gamma} A^{\oplus\to d}_{\gamma\beta} \right|^2,
\end{align}
where we have introduced the flavor transition amplitude from the
source to the Earth surface $A^{s\to \oplus}$ and from the Earth
surface to the detector $A^{\oplus\to d}$.

Generically these amplitudes are obtained by solving the neutrino and
antineutrino evolution equations for the flavor wave function $\vec
\nu(x) = \lbrace \nu_e(x), \nu_\mu(x), \nu_\tau(x) \rbrace^T$
\begin{equation}
  \label{eq:evolvac}
  i\frac{d\vec{\nu}(x)}{dx} = H_\nu^{s\to \oplus} \, \vec{\nu}(x) \,,
  \qquad
  i\frac{d\vec{\bar\nu}(x)}{dx} = H_{\bar\nu}^{s\to \oplus} \, \vec{\bar\nu}(x)
\end{equation}
for evolution between the source and the Earth surface and
\begin{equation}
  \label{eq:evolmat}
  i\frac{d\vec{\nu}(x)}{dx} = H_\nu^{\oplus\to d} \, \vec{\nu}(x) \,,
  \qquad
  i\frac{d\vec{\bar\nu}(x)}{dx} = H_{\bar\nu}^{\oplus\to d}\, \vec{\bar\nu}(x) \,,
\end{equation}
for evolution in the Earth matter.

In this work we are interested in standard vacuum oscillation
dominating the propagation from the source to the detector but
allowing for new physics in the interactions of the neutrinos in the
Earth matter.  In this case
\begin{equation}
  H_\nu^{s\to \oplus} = (H_{\bar\nu}^{s\to \oplus})^*
  = H_\text{osc} = U D_\text{vac} U^\dagger
  \quad\text{with}\quad
  D_\text{vac} = \frac{1}{2 E} \diag(0,\Dmq_{21},\Dmq_{31})
\end{equation}
and $U$ is the leptonic mixing matrix~\cite{Maki:1962mu,
  Kobayashi:1973fv}. While
\begin{equation}
  H_\nu^{\oplus\to d} \simeq H_\text{mat} \,,
  \qquad
  H_{\bar\nu}^{\oplus\to d} \simeq -H^*_\text{mat}
\end{equation}
where the $\simeq$ corresponds to neglecting vacuum oscillations
inside the Earth which is a very good approximation for the relevant
neutrino energies ($\gtrsim 1$ TeV).

The standard theoretical framework for the NP considered here is
provided by non-standard interactions affecting neutrino interactions
in the Earth matter.  They can be described by effective four-fermion
operators of the form
\begin{equation}
  \label{eq:def}
  \mathcal{L}_\text{NSI} =
  - 2\sqrt{2} G_F \Eps_{\alpha\beta}^{fP}
  (\bar\nu_{\alpha} \gamma^\mu \nu_{\beta})
  (\bar{f} \gamma_\mu P f) \,,
\end{equation}
where $f$ is a charged fermion, $P=(L,R)$ and
$\Eps_{\alpha\beta}^{fP}$ are dimensionless parameters encoding the
deviation from standard interactions.  NSI enter in neutrino
propagation only through the vector couplings, so in the most general
case the non-standard matter Hamiltonian can be parametrized
as~\cite{Gonzalez-Garcia:2013usa}
\begin{equation}
  \label{eq:Hmat}
  H_\text{mat} = \sqrt{2} G_F N_e(r)
  \begin{pmatrix}
    1 & 0 & 0 \\
    0 & 0 & 0 \\
    0 & 0 & 0
  \end{pmatrix}
  + \sqrt{2} G_F \sum_{f=e,u,d} N_f(r)
  \begin{pmatrix}
     \Eps_{ee}^f & \Eps_{e\mu}^f & \Eps_{e\tau}^f \\
     \Eps_{e\mu}^{f*} & \Eps_{\mu\mu}^f & \Eps_{\mu\tau}^f \\
     \Eps_{e\tau}^{f*} & \Eps_{\mu\tau}^{f*} & \Eps_{\tau\tau}^f
  \end{pmatrix} \,.
\end{equation}
The standard model interactions are encoded in the non-vanishing $ee$
entry in the first term of Eq.~\eqref{eq:Hmat}, while the non-standard
interactions with fermion $f$ are accounted by the
$\Eps^f_{\alpha\beta}$ coefficients with $\Eps_{\alpha\beta}^f =
\Eps_{\alpha\beta}^{fL} + \Eps_{\alpha\beta}^{fR}$. Here $N_f(r)$ is
the number density of fermions $f$ in the Earth matter.  In practice,
the PREM model~\cite{Dziewonski:1981xy} fixes the neutron/electron
ratio to $Y_n=1.012$ in the Mantle and $Y_n=1.137$ in the Core, with
an average $Y_n=1.051$ all over the Earth.  Thus we get an average
up-quark/electron ratio $Y_u=3.051$ and down-quark/electron ratio
$Y_d=3.102$. We can therefore define:
\begin{equation}
  \Eps_{\alpha\beta}
  \equiv \sum_{f=e,u,d} \left\langle \frac{Y_f}{Y_e} \right\rangle \Eps_{\alpha\beta}^f
  = \Eps_{\alpha\beta}^e + Y_u\, \Eps_{\alpha\beta}^u + Y_d\, \Eps_{\alpha \beta}^d
\end{equation}
so that the matter part of the Hamiltonian can be written as:
\begin{equation}
  H_\text{mat} = \sqrt{2}G_F N_e(r)
  \begin{pmatrix}
    1 + \Eps_{ee} & \Eps_{e\mu} & \Eps_{e\tau} \\
    \Eps_{e\mu}^* & \Eps_{\mu\mu} & \Eps_{\mu\tau} \\
    \Eps_{e\tau}^* & \Eps_{\mu\tau}^* & \Eps_{\tau\tau}
  \end{pmatrix}
  \equiv W D_\text{mat} W^\dagger
\end{equation}
where
\begin{equation}
  D_\text{mat} = \sqrt{2} G_F N_e(r) \diag(\Eps_1,\Eps_2,\Eps_3).
\end{equation}
where $W$ is a $3\times 3$ unitary matrix containing six physical
parameters, three real angles and three complex phases. So without
loss of generality the matter potential contains eight parameters,
five real and three phases (as only difference of $\Eps_i$ enter the
flavor transition probabilities, only differences in the
$\Eps_{\alpha\alpha}$ are physically relevant for neutrino oscillation
data).

\begin{figure}[t]\centering
  \includegraphics[width=0.5\textwidth]{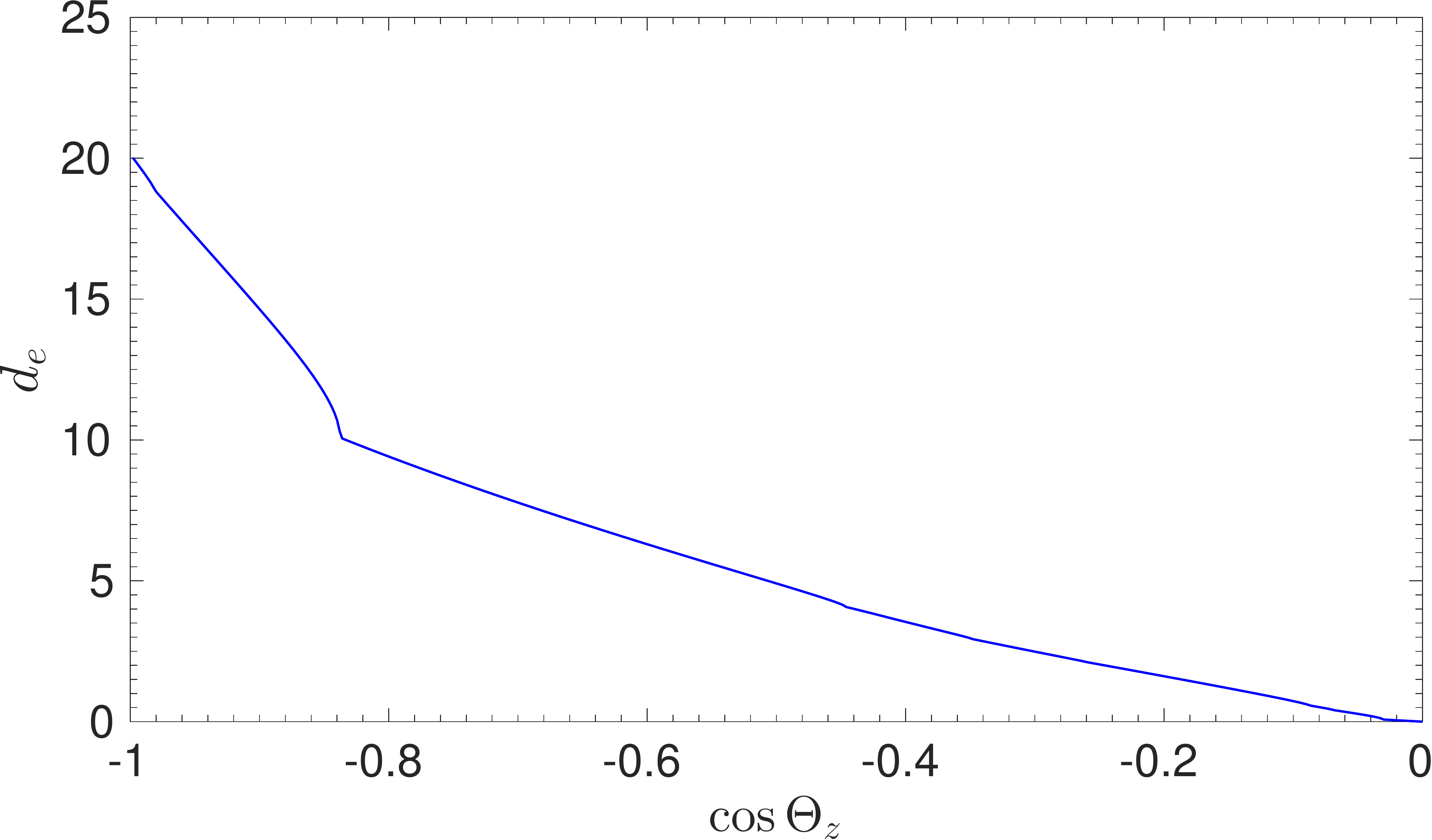}
  \caption{The normalized density integral $d_e$ along the neutrino
    path as a function of the neutrino arrival zenith angle.}
  \label{fig:de}
\end{figure}

Altogether the flavor transition probabilities from a source at
distance $L$ are
\begin{align}
  P^{s\to d}_{\alpha\beta}(E)
  &= \sum_{\gamma\eta k l} W_{\beta k} W_{\beta l}^* W_{\gamma l} W_{\eta k}^*
  \exp(-i d_e\Delta\Eps_{kl})
  \sum_{ij} U_{\eta i}U_{\gamma j}^* U_{\alpha j} U_{\alpha i}^*
  \exp(-i \frac{\Dmq_{ij}}{2E} L) \,,
  \\
  P^{s\to \oplus}_{\alpha\beta}(E)
  &= \sum_{ij} U_{\beta i} U_{\beta j}^* U_{\alpha j} U_{\alpha i}^*
  \exp(-i \frac{\Dmq_{ij}}{2E} L)
\end{align}
where $\Delta \Eps_{kl} = \Eps_k - \Eps_l$. Since for astrophysical
neutrinos the propagation distance $L$ is much longer than the
oscillation wavelength, we can average out the vacuum oscillation
terms:
\begin{align}
  \label{eq:pfin}
  \begin{split}
    P^{s\to d}_{\alpha\beta}(E)
    &= \sum_i |U_{\alpha i}|^2|U_{\beta i}|^2
    - 2 \sum_{\gamma\eta k l i}
    \Re\left(W_{\beta k} W_{\beta l}^* W_{\gamma l} W_{\eta k}^*
    U_{\eta i} U_{\gamma i}^* |U_{\alpha i}|^2 \right)
    \sin^2(d_e \frac{\Delta\Eps_{kl}}{2})
    \\
    & \hphantom{={}} + \sum_{\gamma\eta k l i}
    \Im\left(W_{\beta k} W_{\beta l}^* W_{\gamma l}W_{\eta k}^*
    U_{\eta i} U_{\gamma i}^* |U_{\alpha i}|^2 \right)
    \sin(d_e \Delta\Eps_{kl}) \,,
  \end{split}
  \\
  P^{s\to \oplus}_{\alpha\beta}(E)
  &= \sum_i |U_{\alpha i}|^2|U_{\beta i}|^2 \,.
\end{align}
In these expressions we have introduced the dimensionless
normalization for the matter potential integral along the neutrino
trajectory in the Earth
\begin{equation}
  d_e(\Theta_z) \equiv \int_0^{2R\cos(\pi-\Theta_z)} \sqrt{2}G_FN_e(r) dx \,,
  \quad\text{with}\quad
  r = \sqrt{R_\oplus^2 + x^2 + 2 R_\oplus x \cos\Theta_z} \,,
\end{equation}
which we plot in Fig.~\ref{fig:de}. The integral includes both the
effect of the increase length of the path in the Earth and the
increase average density which is particular relevant for trajectories
crossing the core and leads to the higher slope of the curve for
$\cos\Theta_z \lesssim -0.84$.

We notice that the total coherent flavor transition probability
remains energy independent even in the presence of NSI.  Also the last
term in Eq.~\eqref{eq:pfin} does not change sign for antineutrinos
since both the imaginary part of the combination of mixing matrices
and the phase of the oscillating $\sin$ change sign for
antineutrinos.\footnote{Indeed this term preserves CP but violates
  time reversal, as it is well known that Earth matter effects violate
  CPT.}  In other words, there is no CP violation even if all the
phases in $U$ and $W$ are kept different from zero.  These two facts
render the flavor composition of the fluxes at the detector
independent of the energy spectrum and the neutrino/antineutrino ratio
at the source, as long as the flavor composition at the source is the
same for both neutrinos and antineutrinos. This is just as the case
for standard $3\nu$ oscillations in the absence of
NP.\footnote{Relaxing the assumption of equal flavor composition for
  neutrinos and antineutrinos at the source can lead to additional
  interesting effects even in the case of standard oscillations as
  discussed in Ref.~\cite{Shoemaker:2015qul, Nunokawa:2016pop}.}

In brief, the effect of NSI in the Earth is to modify the flavor
composition at the detector as compared to the standard case, in a way
which depends on the zenith angle of the arrival direction of the
neutrinos. Also, as expected, the effect only appears in presence of
additional flavor mixing during propagation in the Earth,
\textit{i.e.}, for $W_{\alpha i} \neq C \delta_{\alpha i}$, which
occurs only if some off-diagonal $\Eps_{\alpha\beta}$ (with
$\alpha\neq\beta$) is different from zero.

\subsection{Incoherent effects}

In addition to the coherent effects discussed so far, high-energy
neutrinos propagating through the Earth can also interact
inelastically with the Earth matter either by charged current or by
neutral current interactions. As a consequence of these inelastic
processes the neutrino flux is attenuated, its energy is degraded, and
secondary fluxes are generated from the decay of the charged leptons
(in particular $\tau^\pm$) produced in charged current
interactions. In some new physics scenario attenuation and other
decoherence effects can also occur in the travel from the source to
the Earth, but they are not relevant for this work.

For simplicity, let us first neglect NSI and focus only on the usual
$3\nu$ oscillation framework.  In the standard scenario, attenuation
and regeneration effects can be consistently described by a set of
coupled partial integro-differential cascade equations (see for
example~\cite{Jones:2003zy} and references therein).  In this case the
fluxes at the arrival point in the Earth are given by
Eq.~\eqref{eq:eq1} and~\eqref{eq:pse} while for the fluxes at the
detector we have:
\begin{equation}
  \label{eq:fac1}
  \text{SM:}\qquad
  \mathcal{P}_{\alpha\beta}^{s\to d} (E,E') =\sum_{\gamma}
  P^{s\to\oplus}_{\alpha\gamma} (E)
  F^{\oplus\to d}_{\gamma\beta}  (E,E') \,,
\end{equation}
where $F^{\oplus\to d}_{\gamma\beta}(E,E')$ is the function accounting
for attenuation and regeneration effects, which depends on the
trajectory of the neutrino in the Earth matter (\textit{i.e.}, it
depends on $\Theta_z$).  Attenuation is the dominant effect and for
most energies is only mildly flavor dependent. So the dominant
incoherent effects verify
\begin{equation}
  \label{eq:fac2}
  \text{SM:}\qquad
  F^{\oplus\to d}_{\gamma\beta}(E,E')
  \simeq \delta_{\gamma\beta} F_\text{att}^{\oplus\to d}(E) \delta(E-E') \,.
\end{equation}

When considering NSI in the Earth the simple factorization of coherent
and incoherent effects introduced in Eq.~\eqref{eq:fac1} does not
hold, since NSI-induced oscillations, attenuation, and regeneration
occur simultaneously while the neutrino beam is traveling across the
Earth's matter. In order to properly account for all these effects we
need to replace the evolution equation in the Earth~\eqref{eq:evolmat}
with a more general expression including also the incoherent
components. This can be done by means of the density matrix formalism,
as illustrated in Ref.~\cite{GonzalezGarcia:2005xw} (see also
Ref.~\cite{Delgado:2014kpa}).  However, if one neglects the subleading
flavor dependence of these effects and focus only on the dominant
attenuation term, as we did in Eq.~\eqref{eq:fac2} for the standard
case, it becomes possible to write even in the presence of
NSI-oscillations:
\begin{equation}
  \text{NSI:}\qquad
  \mathcal{P}_{\alpha\beta}^{s\to d}(E,E') \simeq
  P_{\alpha\beta}^{s\to d}(E) F_\text{att}^{\oplus\to d}(E) \delta (E-E')
\end{equation}
with $P_{\alpha\beta}^{s\to d}(E)$ given in Eq.~\eqref{eq:pfin}. In
other words, although the presence of NSI affects the flavor
composition at the detector through a modification of the
\emph{coherent} part of the evolution in the Earth, the
\emph{incoherent} part is practically the same in both the standard
and the non-standard case and does not introduce relevant flavor
distortions.

In the next section we quantify our results taking into account the
existing bounds on NSI. For simplicity we will consider only NSI with
quarks and we further assume that the NSI Hamiltonian is real.  At
present the strongest model-independent constraints on NSI with quarks
relevant to neutrino propagation arise from the global analysis of
oscillation data~\cite{Gonzalez-Garcia:2013usa, GonzalezGarcia:2011my}
(see also~\cite{Miranda:2004nb}) in combination with some constraints
from scattering experiments~\cite{Davidson:2003ha, Biggio:2009nt} such
as CHARM~\cite{Dorenbosch:1986tb, Allaby:1987vr},
CDHSW~\cite{Blondel:1989ev} and NuTeV~\cite{Zeller:2001hh}. As shown
in Ref.~\cite{Gonzalez-Garcia:2013usa} neutrino oscillations provide
the stronger constraints on NSI, with the exception of some large
$\Eps_{ee} - \Eps_{\mu\mu}$ terms which are still allowed in
association with a flip of the octant of $\theta_{12}$, the so-called
``dark-side'' solution (or LMA-D) found in
Ref.~\cite{Miranda:2004nb}. However, these large NSI's are disfavored
by scattering data~\cite{Miranda:2004nb}. A fully consistent analysis
of both oscillation and scattering data covering the LMA-D region is
still missing, so here we conservatively consider only NSI's which are
consistent with oscillations within the LMA regions.  The
corresponding allowed ranges read (we quote the most constraining of
both $u$ and $d$ NSI's):
\begin{equation}
  \label{eq:nsiranges}
  \begin{array}{ccc}
    & 90\%~\text{CL} & 3\sigma~\text{CL}
    \\
    \hline
    \Eps_{ee}^q - \Eps_{\mu\mu}^q
    & [+0.02, +0.51]
    & [-0.09, +0.71]
    \\
    \Eps_{\tau\tau}^q - \Eps_{\mu\mu}^q
    & [-0.01, +0.03]
    & [-0.03, +0.19]
    \\
    \Eps_{e\mu}^q
    & [-0.09, +0.04]
    & [-0.16, +0.11]
    \\
    \Eps_{e\tau}^q
    & [-0.13, +0.14]
    & [-0.38, +0.29]
    \\
    \Eps_{\mu\tau}^q
    & [-0.01, +0.01]
    & [-0.03, +0.03]
  \end{array}
\end{equation}
where for each NSI coupling the ranges are shown after marginalization
over all the oscillations parameters and the other NSI couplings.

\section{Results}
\label{sec:results}

Flavor composition of the astrophysical neutrinos are usually
parametrized in terms of the flavor ratios at the source and at the
Earth surface, defined as:
\begin{equation}
  \label{eq:res1}
  \xi^s_\alpha
  \equiv \frac{\phi^s_\alpha(E)}{\sum_\gamma \phi^s_\gamma(E)} \,,
  \qquad
  \xi_\beta^\oplus
  \equiv \frac{\phi^\oplus_\beta(E)}{\sum_\gamma \phi^\oplus_\gamma(E)}
  = \sum_\alpha P^{s\to \oplus}_{\alpha\beta}(E) \xi^s_\alpha
\end{equation}
and it has become customary to plot them in ternary plots.
Experimentally $\xi_\beta^\oplus$ are \emph{reconstructed} from the
measured neutrino fluxes in the detector $\phi_\alpha^d$ by
\emph{deconvoluting} the incoherent effects due to SM interactions in
the Earth matter:
\begin{equation}
  \xi^{\oplus,\text{rec}}_\beta \equiv
  \frac{\phi^{\oplus,\text{rec}}_\beta(E)}{\sum_\gamma \phi^{\oplus,\text{rec}}_\gamma(E)}
  \quad\text{with}\quad
  \phi^{\oplus,\text{rec}}_\beta(E) \equiv
  \sum_{\gamma} \int dE'
  G^{\oplus\leftarrow d}_{\gamma\beta}(E,E') \phi^d_\gamma(E')
\end{equation}
where the function $G^{\oplus\leftarrow d}_{\alpha\beta}(E,E')$ is the
inverse of the Earth $\text{attenuation} + \text{degradation} +
\text{regeneration}$ function $F_{\alpha\beta}^{\oplus\to d}(E,E')$
introduced in the previous section:
\begin{equation}
  \sum_\gamma \int dE'' F^{\oplus\to d}_{\gamma\beta}(E, E'') \,
  G^{\oplus\leftarrow d}_{\alpha\gamma}(E'', E')
  = \delta_{\alpha\beta} \, \delta(E-E')
\end{equation}
Under the approximation described in Eq.~\eqref{eq:fac2}
$G^{\oplus\leftarrow d}_{\alpha\beta}(E,E')$ reduces to:
\begin{equation}
  G^{\oplus\leftarrow d}_{\alpha\beta}(E,E') \simeq
  \delta_{\alpha\beta} \frac{1}{F_\text{att}^{\oplus\to d}(E)} \delta(E-E')
\end{equation}
so that
\begin{equation}
  \xi^{\oplus,\text{rec}}_\beta
  \simeq \frac{\phi^d_\beta(E) \,\big/\, F_\text{att}^{\oplus\to d}(E)}
  {\sum_\gamma \phi^d_\gamma(E) \,\big/\, F_\text{att}^{\oplus\to d}(E)}
  = \frac{\phi^d_\beta(E)}{\sum_\gamma \phi^d_\gamma(E)}
  \equiv \xi^d_\beta
\end{equation}
where we have introduced the flavor ratios at the detector
$\xi^d_\beta$. Thus we have shown that the reconstructed flavor ratios
at the surface of the Earth ($\xi^{\oplus,\text{rec}}_\beta$) are well
approximated by the measured flavor ratios at the detector
($\xi^d_\beta$). This conclusion depends only on the validity of the
approximation~\eqref{eq:fac2}, and therefore applies both for standard
oscillations and in the presence of new physics such as Earth NSI.  It
should be noted, however, that in the standard case
$\xi^{\oplus,\text{rec}}_\beta$ really coincides with the actual
flavor ratios $\xi^\oplus_\beta$ defined in Eq.~\eqref{eq:res1},
whereas in the presence of NSI this is no longer the case.

In what follows we will present our results in terms of flavor ratios
at the detector $\xi^d_\beta$, since, as we have just seen, they are
good estimators of the reconstructed quantities
$\xi^{\oplus,\text{rec}}_\beta$ usually shown by the experimental
collaborations. It is easy to show that:
\begin{equation}
  \xi^d_\beta
  = \sum_\alpha P^{s\to d}_{\alpha\beta}(E) \xi^s_\alpha
\end{equation}
where $P^{s\to d}_{\alpha\beta}(E)$ is obtained from
Eq.~\eqref{eq:pfin}. In principle, one may expect that the flavor
ratios $\xi^d_\beta$ would depend on the neutrino energy, either
through the oscillation probability $P^{s\to d}_{\alpha\beta}(E)$ or
though the intrinsic energy dependence of the flavor ratios at the
source $\xi^s_\alpha$. However, as we have seen in the previous
section the expression in Eq.~\eqref{eq:pfin} is independent of $E$,
and moreover we will assume (as it is customary to do) that the ratios
$\xi^s_\alpha$ do \emph{not} depend on the neutrino energy even though
the fluxes $\phi^s_\alpha(E)$ do. Hence, the flavor ratios
$\xi^d_\beta$ are independent of energy and they can be conveniently
plotted in a ternary plot.

\begin{figure}[t]\centering
  \includegraphics[width=0.3\textwidth]{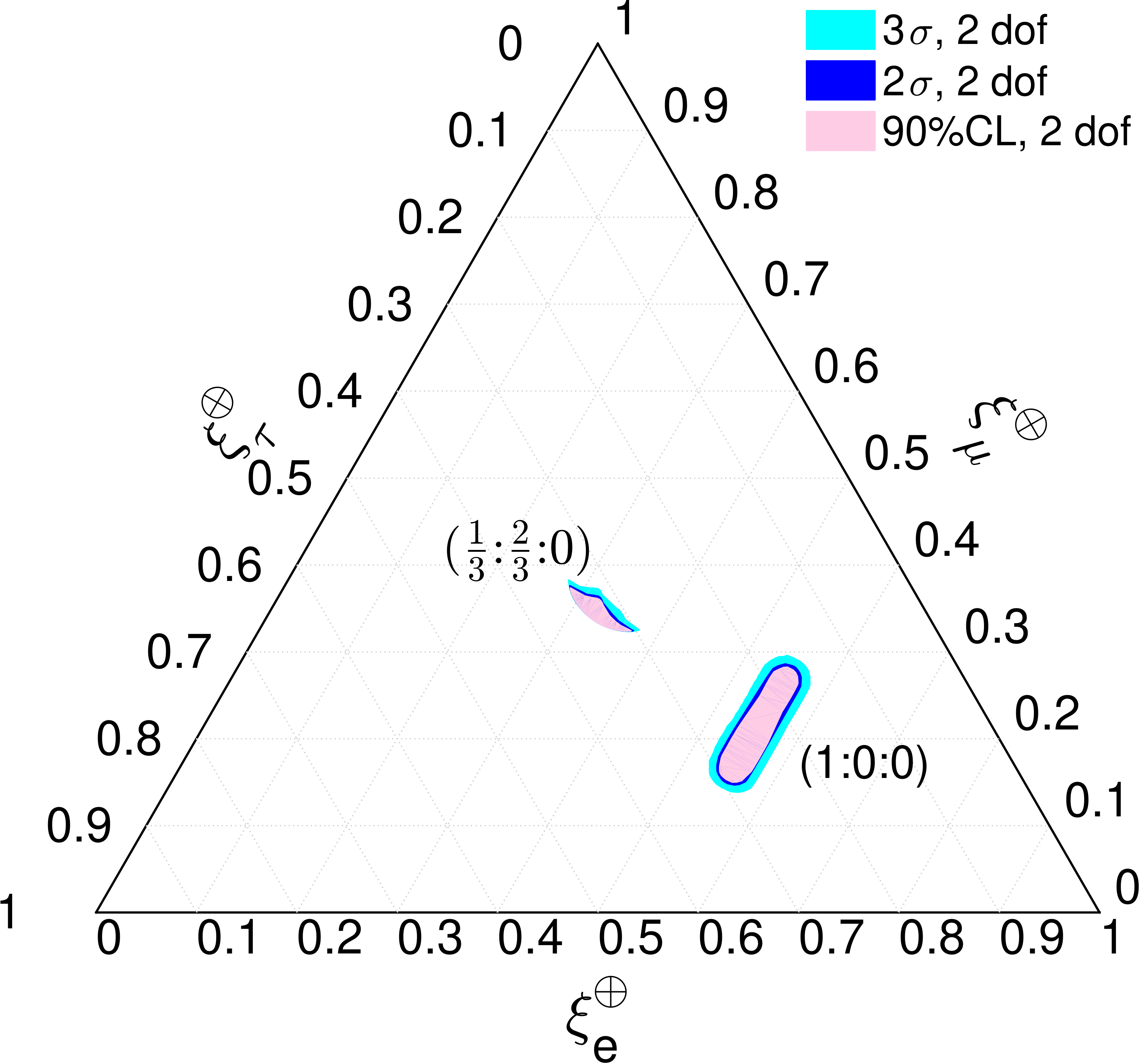}
  \hspace{8mm}
  \includegraphics[width=0.3\textwidth]{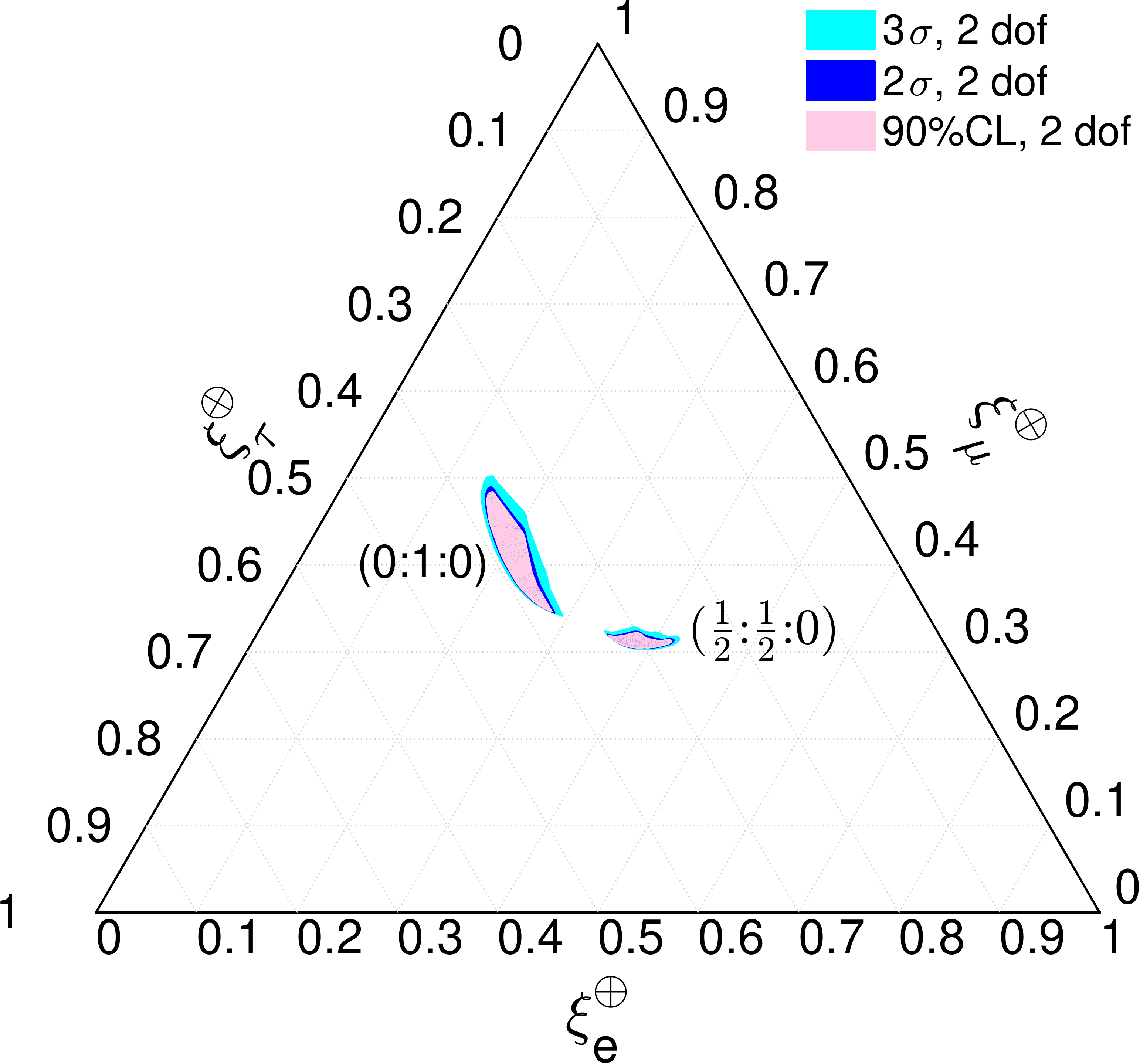}
  \\[5mm]
  \includegraphics[width=0.3\textwidth]{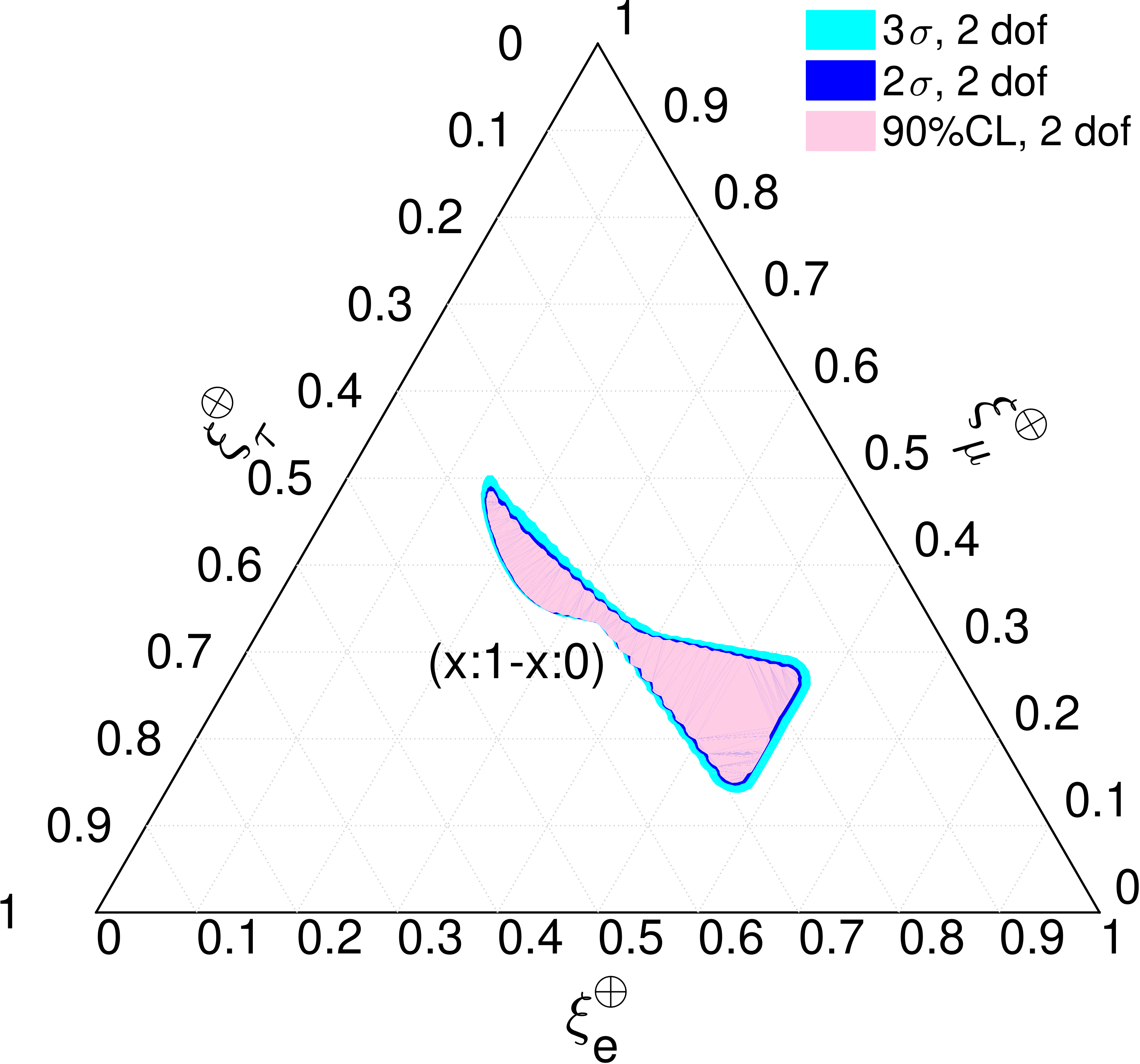}
  \hspace{8mm}
  \includegraphics[width=0.3\textwidth]{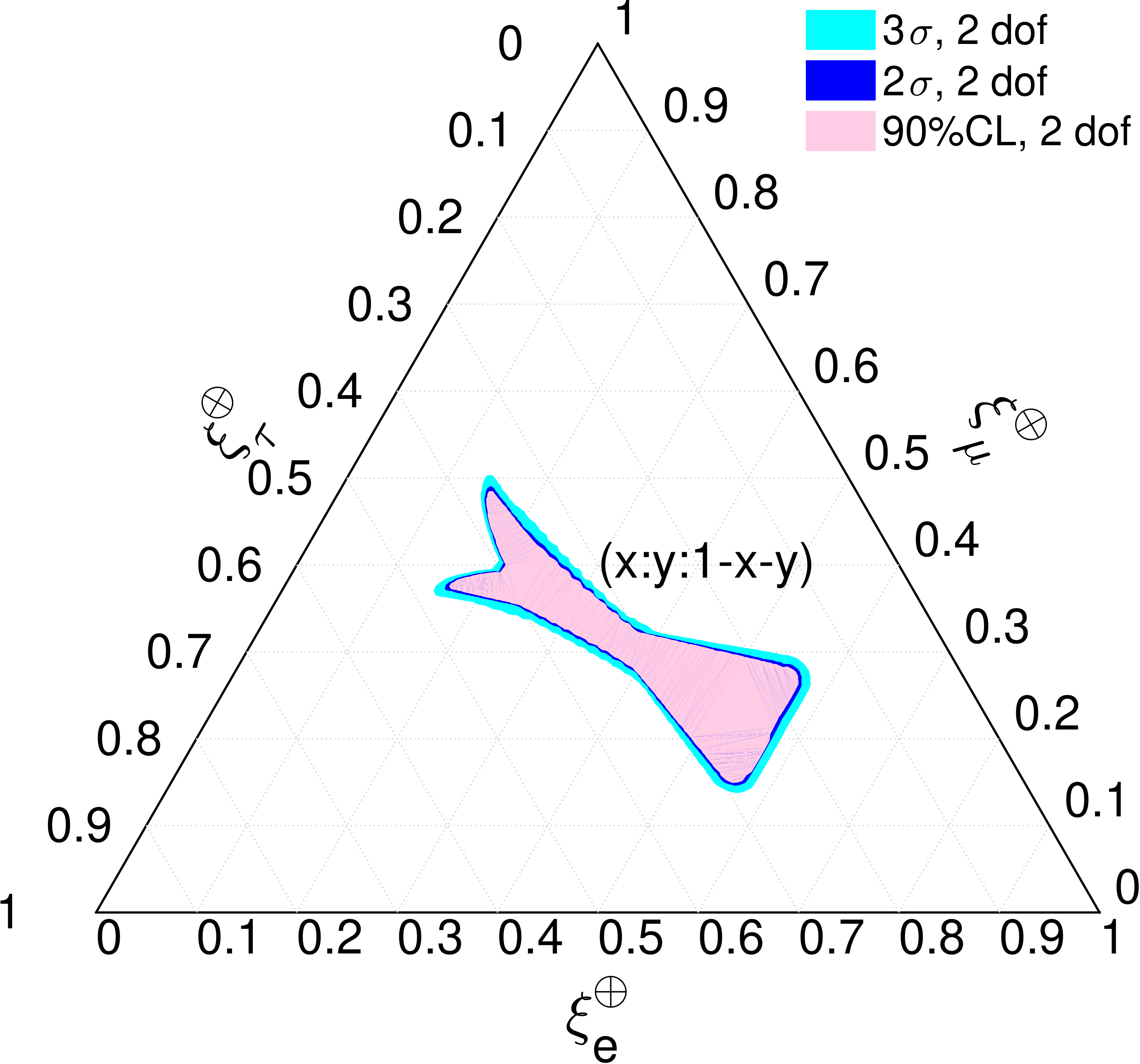}
  \caption{Two-dimensional projections of the allowed regions from the
    global analysis of oscillation data from
    Ref.~\cite{Gonzalez-Garcia:2014fba} in the relevant combinations
    giving the flavor content at the Earth.  The allowed regions are
    shown at 90\%, 95\% and $3\sigma$ CL. In the upper panels we show
    the regions for four initial flavor compositions $(\xi^s_e :
    \xi^s_\mu : \xi^s_\tau) = (\frac{1}{3} : \frac{2}{3} : 0)$,
    $(1:0:0)$, $(0:1:0)$, and $(\frac{1}{2} : \frac{1}{2} : 0)$.  In
    the lower panel the regions are shown for the more general
    scenarios, $(\xi^s_e : \xi^s_\mu : \xi^s_\tau)=(x : 1-x : 0)$ for
    $0\leq x\leq 1$, and $(\xi^s_e:\xi^s_\mu:\xi^s_\tau)=(x : y :
    1-x-y)$ for $0\leq x,y\leq 1$.}
  \label{fig:nufitflav}
\end{figure}

Let us now discuss the results of our fit, starting with the simpler
case of standard oscillations.
In the absence of new physics effects the present determination of the
leptonic mixing matrix from the measurements of neutrino oscillation
experiments allows us to determine the astrophysical neutrino flavor
content at detection given an assumption of the neutrino production
mechanism.  For completeness and reference we show in
Fig.~\ref{fig:nufitflav} the allowed regions of the flavor ratios at
the Earth as obtained from the projection of the six oscillation
parameter $\chi^2$ function of the global NuFIT analysis of
oscillation data~\cite{Gonzalez-Garcia:2014fba, nufit-2.0} in the
relevant mixing combinations (see also~\cite{Arguelles:2015dca,
  Bustamante:2015waa, Nunokawa:2016pop}).  We stress that in our plots
the correlations among the allowed ranges of the oscillation
parameters in the full six-parameter space are properly taken into
account. The results are shown after marginalization over the neutrino
mass ordering and for different assumptions of the flavor content at
the source as labeled in the figure.
Fig.~\ref{fig:nufitflav} illustrates the well-known
fact~\cite{Learned:1994wg} that during propagation from the source
neutrino oscillations lead to flavor content at the Earth close to
$(\xi^\oplus_e : \xi^\oplus_\mu : \xi^\oplus_\tau) = (\frac{1}{3} :
\frac{1}{3} : \frac{1}{3})$, with largest deviations for the case when
the flavor content at the source is $(1:0:0)$~\cite{Palladino:2015vna}
and $(0:1:0$).

\begin{figure}[t]\centering
  \includegraphics[width=0.65\textwidth]{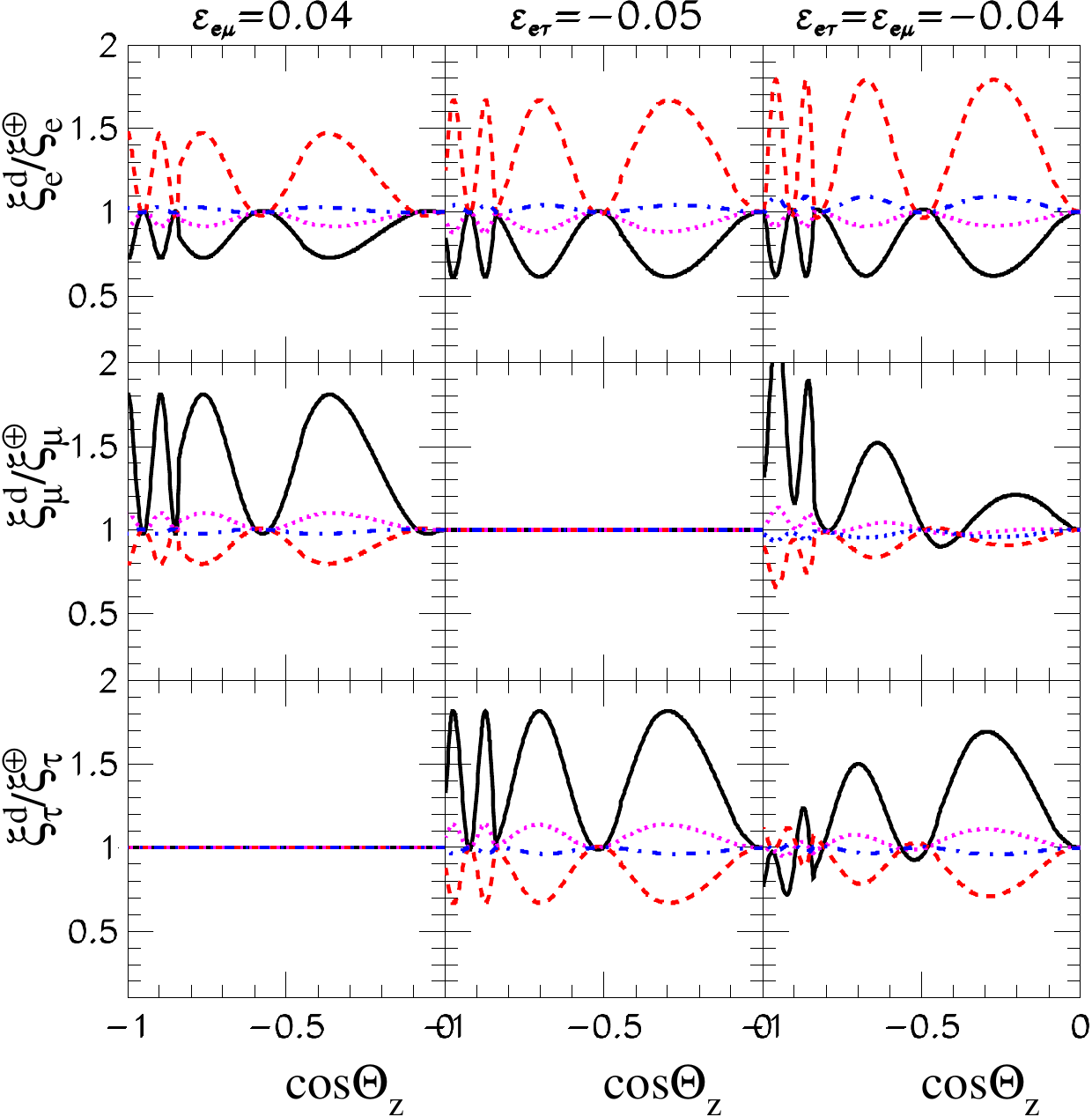}
  \caption{Flavor ratios at the detector as a function of the zenith
    angle of the neutrino normalized to the expectation in the absence
    of NSI and for oscillation parameters at the best fit of the
    global analysis ($\sin^2\theta_{12}= 0.305$, $\sin^2\theta_{13} =
    0.0219$, $\sin^2\theta_{23} = 0.579$, and $\delta_\text{CP} =
    254^\circ$).  For the left (central) [right] panels the only
    non-vanishing NSI parameters are $\Eps_{e\mu}=0.04$ ($\Eps_{e\tau}
    = -0.05$) [$\Eps_{e\mu} = \Eps_{e\tau} = -0.04$]. The different
    curves corresponds to different flavor composition at the source:
    $(\xi^s_e : \xi^s_\mu : \xi^s_\tau)=(1:0:0)$ (full black),
    $(0:1:0)$ (dashed red), $(\frac{1}{2} : \frac{1}{2} : 0)$ (dotted
    blue), and $(\frac{1}{3} : \frac{2}{3} : 0)$ (dash-dotted
    purple).}
  \label{fig:ratios}
\end{figure}

As discussed in the previous section NSI in the Earth modify these
predictions and, unlike for NP effects in the propagation from the
source, such Earth-induced modifications are a function of the arrival
zenith angle of the neutrino. As illustration we show in
Fig.~\ref{fig:ratios} the variation of the flavor ratios at the
detector as a function of the zenith angle of the neutrino for some
values of the $\Eps_{\alpha\beta}$ well within the presently allowed
90\% CL ranges.  In our convention $\cos\Theta_z = -1$ corresponds to
vertically upcoming neutrinos (which have crossed the whole Earth
before reaching the detector) while $\cos\Theta_z = 0$ corresponds to
horizontally arriving neutrinos (for which effectively no Earth matter
is crossed so that $\xi_\beta^d(\cos\Theta_z=0) = \xi_\beta^\oplus$).
From Fig.~\ref{fig:ratios} we can observe the main characteristics of
the effect of NSI in the Earth matter.  Deviations are sizable for
flavor $\alpha$ as long as $\Eps_{\beta\neq\alpha}$ is non-zero and
$\xi^s_\alpha$ or $\xi^s_\beta$ are non-zero. Larger effects are
expected for source flavor compositions for which vacuum oscillations
from the source to the Earth lead to ``less equal'' ratios at the
Earth surface: $(1:0:0)$ and $(0:1:0)$. Finally the increase in
frequency for almost vertical neutrino direction is a consequence of
the increase of the integral density $d_e$ for core crossing
trajectories (see Fig.~\ref{fig:de}).

\begin{figure}[t]\centering
  \includegraphics[height=0.35\textheight]{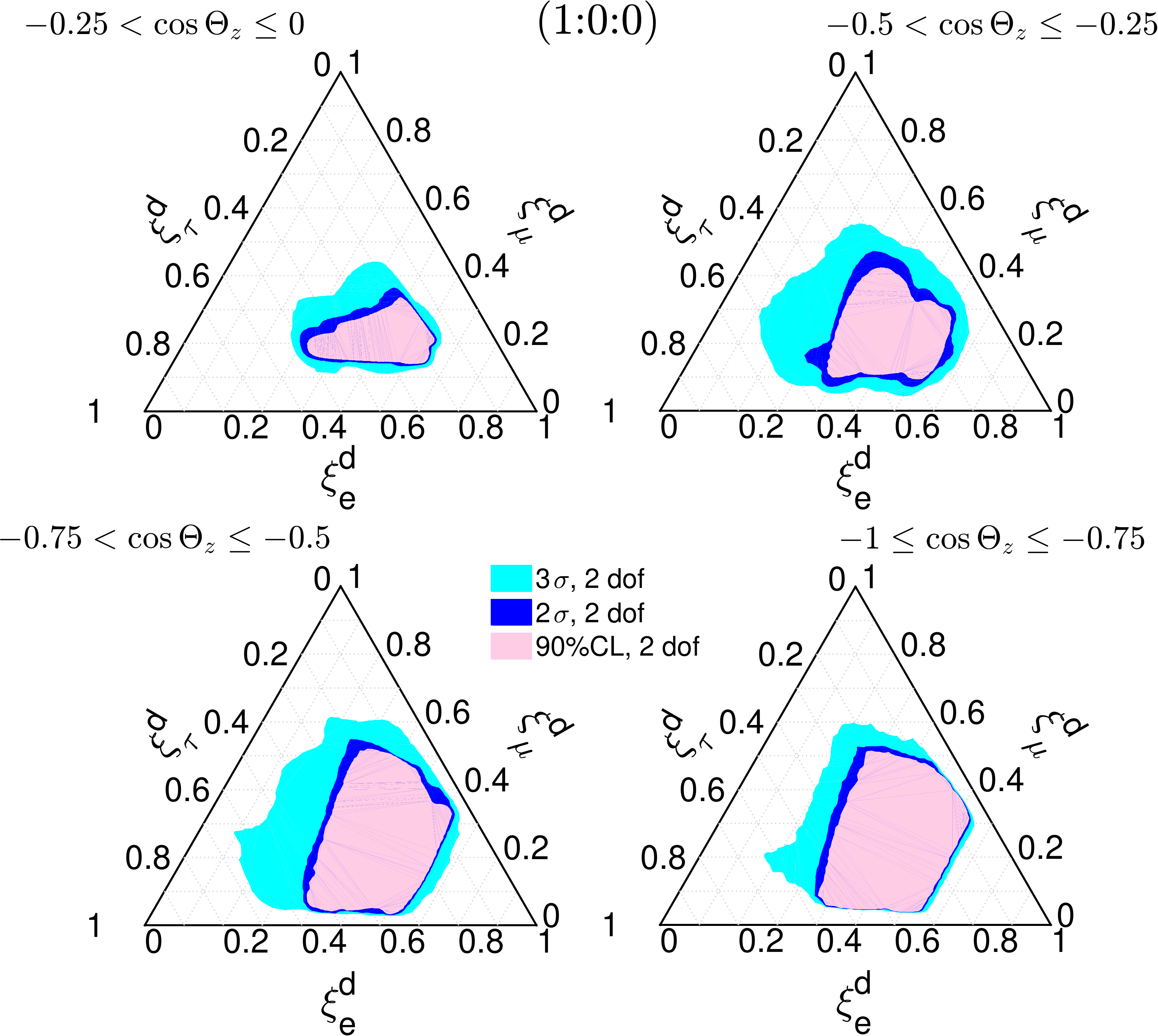}
  \caption{Allowed regions for the flavor ratios in the presence of
    NSI in the Earth at 90, 95\% and 3$\sigma$ CL for an initial
    flavor $(\xi^s_e : \xi^s_\mu : \xi^s_\tau) = (1:0:0)$. The four
    triangles correspond to averaging over neutrinos arriving with
    directions given in the range $0\geq \cos\Theta_z > -0.25$ (upper
    left), $-0.25 \geq \cos\Theta_z > -0.5$ (upper right) $-0.5 \geq
    \cos\Theta_z > -0.75$ (lower left), and $-0.75 \geq \cos\Theta_z
    \geq -1$ (lower right).}
  \label{fig:NSI100}
\end{figure}

\begin{figure}[t]\centering
  \includegraphics[height=0.35\textheight]{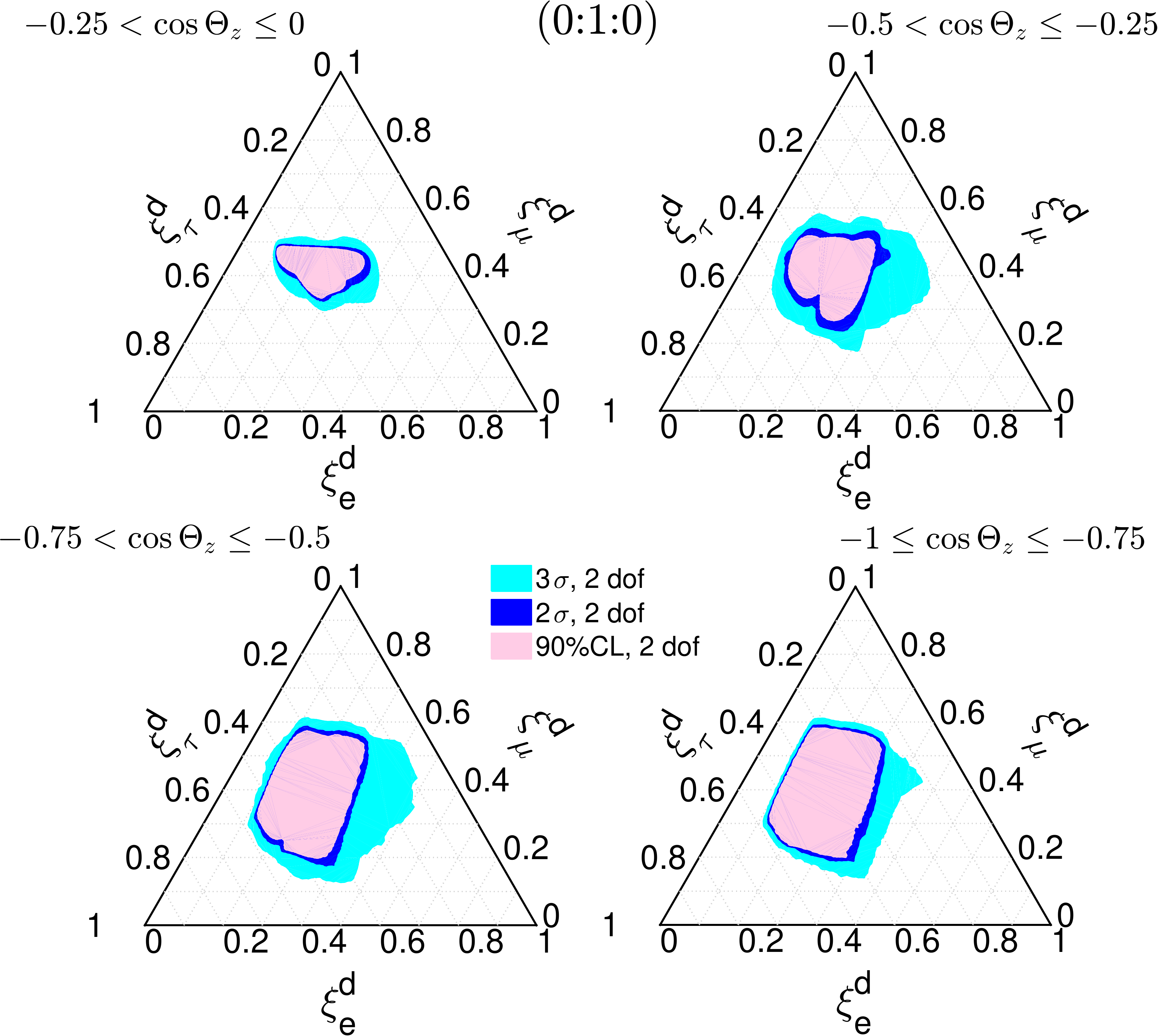}
  \caption{Same as Fig.~\ref{fig:NSI100} for $(\xi^s_e : \xi^s_\mu :
    \xi^s_\tau) = (0:1:0)$.}
  \label{fig:NSI010}
\end{figure}

\begin{figure}[t]\centering
  \includegraphics[height=0.35\textheight]{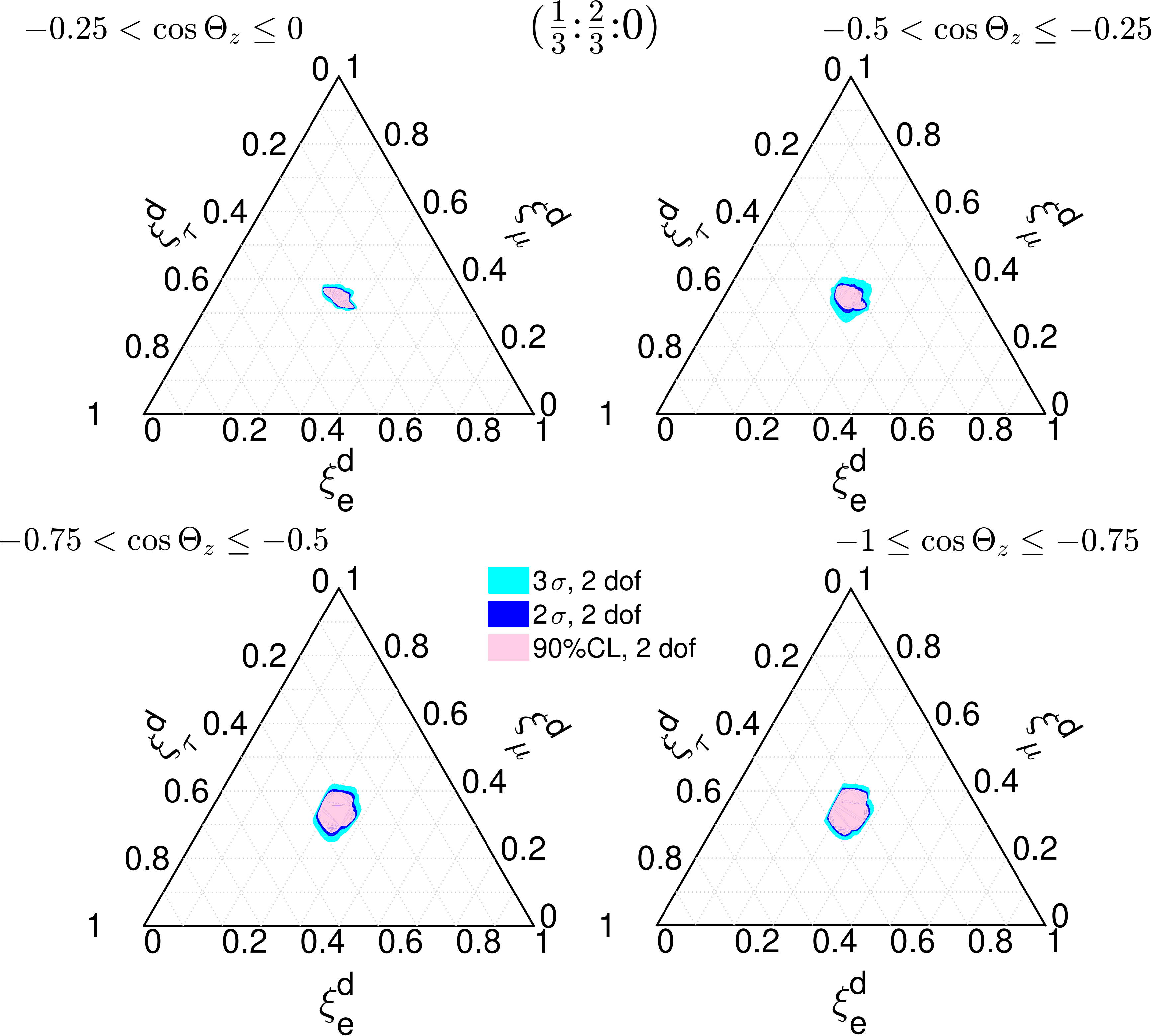}
  \caption{Same as Fig.~\ref{fig:NSI100} for $(\xi^s_e : \xi^s_\mu :
    \xi^s_\tau) = (\frac{1}{3} : \frac{2}{3} : 0)$.}
  \label{fig:NSI120}
\end{figure}

Next we show how the allowed regions in the ternary plots shown in
Fig.~\ref{fig:nufitflav} are modified when including the effect of the
NSI presently allowed at given CL.  In order to do so we project the
$\chi^2$ of the global analysis of oscillation data in the presence of
arbitrary NSI on the relevant combinations entering in the flavor
ratios within a given CL. The results are shown in
Fig.~\ref{fig:NSI100}, Fig.~\ref{fig:NSI010} and Fig.~\ref{fig:NSI120}
for the flavor compositions at source $(\xi^s_e:\xi^s_\mu:\xi^s_\tau)
= (1:0:0)$, $(0:1:0)$ and $(\frac{1}{3}:\frac{2}{3}:0)$,
respectively. The results are shown averaged over four zenith angular
directions.

Comparing the allowed regions in Fig.~\ref{fig:NSI100} and
Fig.~\ref{fig:NSI010} with the corresponding ones for $(1:0:0)$ and
$(0:1:0)$ compositions in the case of standard $3\nu$ oscillations
given in Fig.~\ref{fig:nufitflav} we see that the flavor ratios can
take now much wider range of values in any of the zenith angle ranges
considered. Moreover, although sizable deviations from $(\xi^d_e :
\xi^d_\mu : \xi^d_\tau) = (\frac{1}{3} : \frac{1}{3} : \frac{1}{3})$
are possible, the allowed regions now extend to include $(\frac{1}{3}
: \frac{1}{3} : \frac{1}{3})$ at CL of $3\sigma$ or lower.  We also
see that the larger CL region becomes smaller for most vertical
arrival directions (see the relative size of the light blue regions in
the two lower triangles on these figures). This is so because at those
CL for the larger values of $\Eps$ allowed the NSI-induced
oscillations are fast enough to be averaged out $\langle
\sin^2(\Delta\Eps_{ij} \frac{d_e}{2})\rangle \sim \frac{1}{2}$ for
those trajectories while the value in the second most vertical angular
bin can be in average larger than $1/2$.
For contrast, as illustrated in Fig.~\ref{fig:NSI120}, for the case of
flavor composition at the source $(\xi^s_e : \xi^s_\mu : \xi^s_\tau) =
(\frac{1}{3} : \frac{2}{3} : 0)$ NSI in the Earth never induce sizable
modifications of the expectation $(\xi^d_e : \xi^d_\mu : \xi^d_\tau)=
(\frac{1}{3} : \frac{1}{3} : \frac{1}{3})$.

\section{Conclusions}
\label{sec:conclu}

The measurement of the flavor composition of the detected ultra-high
energy neutrinos can be a powerful tool to learn about the mechanisms
at work in their sources. Such inference, however, relies on the
understanding of the particle physics processes relevant to the
neutrino propagation from the source to the detector. The presence of
NP effects beyond those of the well established mass-induced $3\nu$
oscillations alter the flavor composition at the detector and can
therefore affect the conclusions on the dominant production mechanism.

In this work we have focused on NP effects associated with NSI of the
neutrinos in the Earth matter. The relevant flavor transition
probabilities accounting from oscillations from the source to the
Earth plus NSI in the Earth are energy independent but depend on the
zenith angle of the arrival direction of the neutrinos, which is a
characteristic feature of this form of NP.  Quantitatively, we have
shown that within the presently allowed range of NSI large deviations
from the standard $3\nu$ oscillation predictions for the detected
flavor composition can be expected, in particular for fluxes dominated
by one flavor at the source. On the contrary we find that the
expectation of equalized flavors in the Earth for sources dominated by
production via pion-muon decay-chain is robust even in the presence of
this form of NP.

\section*{Acknowledgments}

I.M.S.\ thanks YITP at Stony Brook Univ.\ for their kind hospitality
during the visit that lead to this work.  This work is supported by
USA-NSF grant PHY-13-16617, by EU Networks FP7 ITN INVISIBLES
(PITN-GA-2011-289442), FP10 ITN ELUSIVES (H2020-MSCA-ITN-2015-674896)
and INVISIBLES-PLUS (H2020-MSCA-RISE-2015-690575).  M.C.G-G.\ also
acknowledges support by MINECO grants 2014-SGR-104, FPA2013-46570, and
``Maria de Maetzu'' program grant MDM-2014-0367 of ICCUB.  M.M.\ and
I.M-S.\ also acknowledge support by MINECO grants FPA2012-31880,
FPA2012-34694 and by the ``Severo Ochoa'' program grant SEV-2012-0249
of IFT.

\section*{References}

\bibliographystyle{elsarticle-num}
\bibliography{biblio}

\begin{thebibliography}{10}
\expandafter\ifx\csname url\endcsname\relax
  \def\url#1{\texttt{#1}}\fi
\expandafter\ifx\csname urlprefix\endcsname\relax\def\urlprefix{URL }\fi
\expandafter\ifx\csname href\endcsname\relax
  \def\href#1#2{#2} \def\path#1{#1}\fi

\bibitem{Aartsen:2013bka}
M.~G. Aartsen, et~al., {First observation of PeV-energy neutrinos with
  IceCube}, Phys. Rev. Lett. 111 (2013) 021103.
\newblock \href {http://arxiv.org/abs/1304.5356} {\path{arXiv:1304.5356}},
  \href {http://dx.doi.org/10.1103/PhysRevLett.111.021103}
  {\path{doi:10.1103/PhysRevLett.111.021103}}.

\bibitem{Aartsen:2013jdh}
M.~G. Aartsen, et~al., {Evidence for High-Energy Extraterrestrial Neutrinos at
  the IceCube Detector}, Science 342 (2013) 1242856.
\newblock \href {http://arxiv.org/abs/1311.5238} {\path{arXiv:1311.5238}},
  \href {http://dx.doi.org/10.1126/science.1242856}
  {\path{doi:10.1126/science.1242856}}.

\bibitem{Aartsen:2014gkd}
M.~G. Aartsen, et~al., {Observation of High-Energy Astrophysical Neutrinos in
  Three Years of IceCube Data}, Phys. Rev. Lett. 113 (2014) 101101.
\newblock \href {http://arxiv.org/abs/1405.5303} {\path{arXiv:1405.5303}},
  \href {http://dx.doi.org/10.1103/PhysRevLett.113.101101}
  {\path{doi:10.1103/PhysRevLett.113.101101}}.

\bibitem{Aartsen:2015rwa}
M.~G. Aartsen, et~al., {Evidence for Astrophysical Muon Neutrinos from the
  Northern Sky with IceCube}, Phys. Rev. Lett. 115~(8) (2015) 081102.
\newblock \href {http://arxiv.org/abs/1507.04005} {\path{arXiv:1507.04005}},
  \href {http://dx.doi.org/10.1103/PhysRevLett.115.081102}
  {\path{doi:10.1103/PhysRevLett.115.081102}}.

\bibitem{Anchordoqui:2013dnh}
L.~A. Anchordoqui, et~al., {Cosmic Neutrino Pevatrons: A Brand New Pathway to
  Astronomy, Astrophysics, and Particle Physics}, JHEAp 1-2 (2014) 1--30.
\newblock \href {http://arxiv.org/abs/1312.6587} {\path{arXiv:1312.6587}},
  \href {http://dx.doi.org/10.1016/j.jheap.2014.01.001}
  {\path{doi:10.1016/j.jheap.2014.01.001}}.

\bibitem{Learned:1994wg}
J.~G. Learned, S.~Pakvasa, {Detecting tau-neutrino oscillations at PeV
  energies}, Astropart. Phys. 3 (1995) 267--274.
\newblock \href {http://arxiv.org/abs/hep-ph/9405296}
  {\path{arXiv:hep-ph/9405296}}, \href
  {http://dx.doi.org/10.1016/0927-6505(94)00043-3}
  {\path{doi:10.1016/0927-6505(94)00043-3}}.

\bibitem{Kashti:2005qa}
T.~Kashti, E.~Waxman, {Flavoring astrophysical neutrinos: Flavor ratios depend
  on energy}, Phys. Rev. Lett. 95 (2005) 181101.
\newblock \href {http://arxiv.org/abs/astro-ph/0507599}
  {\path{arXiv:astro-ph/0507599}}, \href
  {http://dx.doi.org/10.1103/PhysRevLett.95.181101}
  {\path{doi:10.1103/PhysRevLett.95.181101}}.

\bibitem{Lipari:2007su}
P.~Lipari, M.~Lusignoli, D.~Meloni, {Flavor Composition and Energy Spectrum of
  Astrophysical Neutrinos}, Phys. Rev. D75 (2007) 123005.
\newblock \href {http://arxiv.org/abs/0704.0718} {\path{arXiv:0704.0718}},
  \href {http://dx.doi.org/10.1103/PhysRevD.75.123005}
  {\path{doi:10.1103/PhysRevD.75.123005}}.

\bibitem{Kachelriess:2007tr}
M.~Kachelriess, S.~Ostapchenko, R.~Tomas, {High energy neutrino yields from
  astrophysical sources. 2. Magnetized sources}, Phys. Rev. D77 (2008) 023007.
\newblock \href {http://arxiv.org/abs/0708.3047} {\path{arXiv:0708.3047}},
  \href {http://dx.doi.org/10.1103/PhysRevD.77.023007}
  {\path{doi:10.1103/PhysRevD.77.023007}}.

\bibitem{Hummer:2010ai}
S.~Hummer, M.~Maltoni, W.~Winter, C.~Yaguna, {Energy dependent neutrino flavor
  ratios from cosmic accelerators on the Hillas plot}, Astropart. Phys. 34
  (2010) 205--224.
\newblock \href {http://arxiv.org/abs/1007.0006} {\path{arXiv:1007.0006}},
  \href {http://dx.doi.org/10.1016/j.astropartphys.2010.07.003}
  {\path{doi:10.1016/j.astropartphys.2010.07.003}}.

\bibitem{Winter:2014pya}
W.~Winter, {Describing the Observed Cosmic Neutrinos by Interactions of Nuclei
  with Matter}, Phys. Rev. D90~(10) (2014) 103003.
\newblock \href {http://arxiv.org/abs/1407.7536} {\path{arXiv:1407.7536}},
  \href {http://dx.doi.org/10.1103/PhysRevD.90.103003}
  {\path{doi:10.1103/PhysRevD.90.103003}}.

\bibitem{Lunardini:2000fy}
C.~Lunardini, A.~{\relax Yu}. Smirnov, {High-energy neutrino conversion and the
  lepton asymmetry in the universe}, Phys. Rev. D64 (2001) 073006.
\newblock \href {http://arxiv.org/abs/hep-ph/0012056}
  {\path{arXiv:hep-ph/0012056}}, \href
  {http://dx.doi.org/10.1103/PhysRevD.64.073006}
  {\path{doi:10.1103/PhysRevD.64.073006}}.

\bibitem{Razzaque:2009kq}
S.~Razzaque, A.~{\relax Yu}. Smirnov, {Flavor conversion of cosmic neutrinos
  from hidden jets}, JHEP 03 (2010) 031.
\newblock \href {http://arxiv.org/abs/0912.4028} {\path{arXiv:0912.4028}},
  \href {http://dx.doi.org/10.1007/JHEP03(2010)031}
  {\path{doi:10.1007/JHEP03(2010)031}}.

\bibitem{Sahu:2010ap}
S.~Sahu, B.~Zhang, {Effect of Resonant Neutrino Oscillation on TeV Neutrino
  Flavor Ratio from Choked GRBs}, Res. Astron. Astrophys. 10 (2010) 943--949.
\newblock \href {http://arxiv.org/abs/1007.4582} {\path{arXiv:1007.4582}},
  \href {http://dx.doi.org/10.1088/1674-4527/10/10/001}
  {\path{doi:10.1088/1674-4527/10/10/001}}.

\bibitem{Mena:2014sja}
O.~Mena, S.~Palomares-Ruiz, A.~C. Vincent, {Flavor Composition of the
  High-Energy Neutrino Events in IceCube}, Phys. Rev. Lett. 113 (2014) 091103.
\newblock \href {http://arxiv.org/abs/1404.0017} {\path{arXiv:1404.0017}},
  \href {http://dx.doi.org/10.1103/PhysRevLett.113.091103}
  {\path{doi:10.1103/PhysRevLett.113.091103}}.

\bibitem{Bustamante:2015waa}
M.~Bustamante, J.~F. Beacom, W.~Winter, {Theoretically palatable flavor
  combinations of astrophysical neutrinos}, Phys. Rev. Lett. 115~(16) (2015)
  161302.
\newblock \href {http://arxiv.org/abs/1506.02645} {\path{arXiv:1506.02645}},
  \href {http://dx.doi.org/10.1103/PhysRevLett.115.161302}
  {\path{doi:10.1103/PhysRevLett.115.161302}}.

\bibitem{Palladino:2015zua}
A.~Palladino, G.~Pagliaroli, F.~Villante, F.~Vissani, {What is the Flavor of
  the Cosmic Neutrinos Seen by IceCube?}, Phys. Rev. Lett. 114~(17) (2015)
  171101.
\newblock \href {http://arxiv.org/abs/1502.02923} {\path{arXiv:1502.02923}},
  \href {http://dx.doi.org/10.1103/PhysRevLett.114.171101}
  {\path{doi:10.1103/PhysRevLett.114.171101}}.

\bibitem{Palomares-Ruiz:2015mka}
S.~Palomares-Ruiz, A.~C. Vincent, O.~Mena, {Spectral analysis of the
  high-energy IceCube neutrinos}, Phys. Rev. D91~(10) (2015) 103008.
\newblock \href {http://arxiv.org/abs/1502.02649} {\path{arXiv:1502.02649}},
  \href {http://dx.doi.org/10.1103/PhysRevD.91.103008}
  {\path{doi:10.1103/PhysRevD.91.103008}}.

\bibitem{Watanabe:2014qua}
A.~Watanabe, {The spectrum and flavor composition of the astrophysical
  neutrinos in IceCube}, JCAP 1508 (2015) 030.
\newblock \href {http://arxiv.org/abs/1412.8264} {\path{arXiv:1412.8264}},
  \href {http://dx.doi.org/10.1088/1475-7516/2015/08/030}
  {\path{doi:10.1088/1475-7516/2015/08/030}}.

\bibitem{Kawanaka:2015qza}
N.~Kawanaka, K.~Ioka, {Neutrino Flavor Ratios Modified by Cosmic Ray Secondary
  Acceleration}, Phys. Rev. D92~(8) (2015) 085047.
\newblock \href {http://arxiv.org/abs/1504.03417} {\path{arXiv:1504.03417}},
  \href {http://dx.doi.org/10.1103/PhysRevD.92.085047}
  {\path{doi:10.1103/PhysRevD.92.085047}}.

\bibitem{Aartsen:2015knd}
M.~G. Aartsen, et~al., {A combined maximum-likelihood analysis of the
  high-energy astrophysical neutrino flux measured with IceCube}, Astrophys. J.
  809~(1) (2015) 98.
\newblock \href {http://arxiv.org/abs/1507.03991} {\path{arXiv:1507.03991}},
  \href {http://dx.doi.org/10.1088/0004-637X/809/1/98}
  {\path{doi:10.1088/0004-637X/809/1/98}}.

\bibitem{Vincent:2016nut}
A.~C. Vincent, S.~Palomares-Ruiz, O.~Mena, {Analysis of the 4-year IceCube HESE
  data}\href {http://arxiv.org/abs/1605.01556} {\path{arXiv:1605.01556}}.

\bibitem{Hooper:2005jp}
D.~Hooper, D.~Morgan, E.~Winstanley, {Lorentz and CPT invariance violation in
  high-energy neutrinos}, Phys. Rev. D72 (2005) 065009.
\newblock \href {http://arxiv.org/abs/hep-ph/0506091}
  {\path{arXiv:hep-ph/0506091}}, \href
  {http://dx.doi.org/10.1103/PhysRevD.72.065009}
  {\path{doi:10.1103/PhysRevD.72.065009}}.

\bibitem{Beacom:2002vi}
J.~F. Beacom, N.~F. Bell, D.~Hooper, S.~Pakvasa, T.~J. Weiler, {Decay of
  high-energy astrophysical neutrinos}, Phys. Rev. Lett. 90 (2003) 181301.
\newblock \href {http://arxiv.org/abs/hep-ph/0211305}
  {\path{arXiv:hep-ph/0211305}}, \href
  {http://dx.doi.org/10.1103/PhysRevLett.90.181301}
  {\path{doi:10.1103/PhysRevLett.90.181301}}.

\bibitem{Baerwald:2012kc}
P.~Baerwald, M.~Bustamante, W.~Winter, {Neutrino Decays over Cosmological
  Distances and the Implications for Neutrino Telescopes}, JCAP 1210 (2012)
  020.
\newblock \href {http://arxiv.org/abs/1208.4600} {\path{arXiv:1208.4600}},
  \href {http://dx.doi.org/10.1088/1475-7516/2012/10/020}
  {\path{doi:10.1088/1475-7516/2012/10/020}}.

\bibitem{Anchordoqui:2005gj}
L.~A. Anchordoqui, H.~Goldberg, M.~C. Gonzalez-Garcia, F.~Halzen, D.~Hooper,
  S.~Sarkar, T.~J. Weiler, {Probing Planck scale physics with IceCube}, Phys.
  Rev. D72 (2005) 065019.
\newblock \href {http://arxiv.org/abs/hep-ph/0506168}
  {\path{arXiv:hep-ph/0506168}}, \href
  {http://dx.doi.org/10.1103/PhysRevD.72.065019}
  {\path{doi:10.1103/PhysRevD.72.065019}}.

\bibitem{Hooper:2004xr}
D.~Hooper, D.~Morgan, E.~Winstanley, {Probing quantum decoherence with
  high-energy neutrinos}, Phys. Lett. B609 (2005) 206--211.
\newblock \href {http://arxiv.org/abs/hep-ph/0410094}
  {\path{arXiv:hep-ph/0410094}}, \href
  {http://dx.doi.org/10.1016/j.physletb.2005.01.034}
  {\path{doi:10.1016/j.physletb.2005.01.034}}.

\bibitem{Beacom:2003eu}
J.~F. Beacom, N.~F. Bell, D.~Hooper, J.~G. Learned, S.~Pakvasa, T.~J. Weiler,
  {PseudoDirac neutrinos: A Challenge for neutrino telescopes}, Phys. Rev.
  Lett. 92 (2004) 011101.
\newblock \href {http://arxiv.org/abs/hep-ph/0307151}
  {\path{arXiv:hep-ph/0307151}}, \href
  {http://dx.doi.org/10.1103/PhysRevLett.92.011101}
  {\path{doi:10.1103/PhysRevLett.92.011101}}.

\bibitem{Esmaili:2009fk}
A.~Esmaili, {Pseudo-Dirac Neutrino Scenario: Cosmic Neutrinos at Neutrino
  Telescopes}, Phys. Rev. D81 (2010) 013006.
\newblock \href {http://arxiv.org/abs/0909.5410} {\path{arXiv:0909.5410}},
  \href {http://dx.doi.org/10.1103/PhysRevD.81.013006}
  {\path{doi:10.1103/PhysRevD.81.013006}}.

\bibitem{Athar:2000yw}
H.~Athar, M.~Jezabek, O.~Yasuda, {Effects of neutrino mixing on high-energy
  cosmic neutrino flux}, Phys. Rev. D62 (2000) 103007.
\newblock \href {http://arxiv.org/abs/hep-ph/0005104}
  {\path{arXiv:hep-ph/0005104}}, \href
  {http://dx.doi.org/10.1103/PhysRevD.62.103007}
  {\path{doi:10.1103/PhysRevD.62.103007}}.

\bibitem{deSalas:2016svi}
P.~F. de~Salas, R.~A. Lineros, M.~Tórtola, {Neutrino propagation in the
  galactic dark matter halo}\href {http://arxiv.org/abs/1601.05798}
  {\path{arXiv:1601.05798}}.

\bibitem{Arguelles:2015dca}
C.~A. Argüelles, T.~Katori, J.~Salvado, {New Physics in Astrophysical Neutrino
  Flavor}, Phys. Rev. Lett. 115 (2015) 161303.
\newblock \href {http://arxiv.org/abs/1506.02043} {\path{arXiv:1506.02043}},
  \href {http://dx.doi.org/10.1103/PhysRevLett.115.161303}
  {\path{doi:10.1103/PhysRevLett.115.161303}}.

\bibitem{Maki:1962mu}
Z.~Maki, M.~Nakagawa, S.~Sakata, {Remarks on the unified model of elementary
  particles}, Prog. Theor. Phys. 28 (1962) 870--880.
\newblock \href {http://dx.doi.org/10.1143/PTP.28.870}
  {\path{doi:10.1143/PTP.28.870}}.

\bibitem{Kobayashi:1973fv}
M.~Kobayashi, T.~Maskawa, {CP Violation in the Renormalizable Theory of Weak
  Interaction}, Prog. Theor. Phys. 49 (1973) 652--657.
\newblock \href {http://dx.doi.org/10.1143/PTP.49.652}
  {\path{doi:10.1143/PTP.49.652}}.

\bibitem{Gonzalez-Garcia:2013usa}
M.~C. Gonzalez-Garcia, M.~Maltoni, {Determination of matter potential from
  global analysis of neutrino oscillation data}, JHEP 09 (2013) 152.
\newblock \href {http://arxiv.org/abs/1307.3092} {\path{arXiv:1307.3092}},
  \href {http://dx.doi.org/10.1007/JHEP09(2013)152}
  {\path{doi:10.1007/JHEP09(2013)152}}.

\bibitem{Dziewonski:1981xy}
A.~Dziewonski, D.~Anderson, {Preliminary reference earth model}, Phys.Earth
  Planet.Interiors 25 (1981) 297--356.
\newblock \href {http://dx.doi.org/10.1016/0031-9201(81)90046-7}
  {\path{doi:10.1016/0031-9201(81)90046-7}}.

\bibitem{Shoemaker:2015qul}
I.~M. Shoemaker, K.~Murase, {Probing BSM Neutrino Physics with Flavor and
  Spectral Distortions: Prospects for Future High-Energy Neutrino Telescopes},
  Phys. Rev. D93~(8) (2016) 085004.
\newblock \href {http://arxiv.org/abs/1512.07228} {\path{arXiv:1512.07228}},
  \href {http://dx.doi.org/10.1103/PhysRevD.93.085004}
  {\path{doi:10.1103/PhysRevD.93.085004}}.

\bibitem{Nunokawa:2016pop}
H.~Nunokawa, B.~Panes, R.~Z. Funchal, {How Unequal Fluxes of High Energy
  Astrophysical Neutrinos and Antineutrinos can Fake New Physics}\href
  {http://arxiv.org/abs/1604.08595} {\path{arXiv:1604.08595}}.

\bibitem{Jones:2003zy}
J.~Jones, I.~Mocioiu, M.~H. Reno, I.~Sarcevic, {Tracing very high-energy
  neutrinos from cosmological distances in ice}, Phys. Rev. D69 (2004) 033004.
\newblock \href {http://arxiv.org/abs/hep-ph/0308042}
  {\path{arXiv:hep-ph/0308042}}, \href
  {http://dx.doi.org/10.1103/PhysRevD.69.033004}
  {\path{doi:10.1103/PhysRevD.69.033004}}.

\bibitem{GonzalezGarcia:2005xw}
M.~C. Gonzalez-Garcia, F.~Halzen, M.~Maltoni, {Physics reach of high-energy and
  high-statistics icecube atmospheric neutrino data}, Phys. Rev. D71 (2005)
  093010.
\newblock \href {http://arxiv.org/abs/hep-ph/0502223}
  {\path{arXiv:hep-ph/0502223}}, \href
  {http://dx.doi.org/10.1103/PhysRevD.71.093010}
  {\path{doi:10.1103/PhysRevD.71.093010}}.

\bibitem{Delgado:2014kpa}
C.~A. Argüelles~Delgado, J.~Salvado, C.~N. Weaver, {A Simple Quantum
  Integro-Differential Solver (SQuIDS)}, Comput. Phys. Commun. 196 (2015)
  569--591.
\newblock \href {http://arxiv.org/abs/1412.3832} {\path{arXiv:1412.3832}},
  \href {http://dx.doi.org/10.1016/j.cpc.2015.06.022}
  {\path{doi:10.1016/j.cpc.2015.06.022}}.

\bibitem{GonzalezGarcia:2011my}
M.~C. Gonzalez-Garcia, M.~Maltoni, J.~Salvado, {Testing matter effects in
  propagation of atmospheric and long-baseline neutrinos}, JHEP 05 (2011) 075.
\newblock \href {http://arxiv.org/abs/1103.4365} {\path{arXiv:1103.4365}},
  \href {http://dx.doi.org/10.1007/JHEP05(2011)075}
  {\path{doi:10.1007/JHEP05(2011)075}}.

\bibitem{Miranda:2004nb}
O.~Miranda, M.~Tortola, J.~Valle, {Are solar neutrino oscillations robust?},
  JHEP 0610 (2006) 008.
\newblock \href {http://arxiv.org/abs/hep-ph/0406280}
  {\path{arXiv:hep-ph/0406280}}, \href
  {http://dx.doi.org/10.1088/1126-6708/2006/10/008}
  {\path{doi:10.1088/1126-6708/2006/10/008}}.

\bibitem{Davidson:2003ha}
S.~Davidson, C.~Pena-Garay, N.~Rius, A.~Santamaria, {Present and future bounds
  on nonstandard neutrino interactions}, JHEP 0303 (2003) 011.
\newblock \href {http://arxiv.org/abs/hep-ph/0302093}
  {\path{arXiv:hep-ph/0302093}}.

\bibitem{Biggio:2009nt}
C.~Biggio, M.~Blennow, E.~Fernandez-Martinez, {General bounds on non-standard
  neutrino interactions}, JHEP 0908 (2009) 090.
\newblock \href {http://arxiv.org/abs/0907.0097} {\path{arXiv:0907.0097}},
  \href {http://dx.doi.org/10.1088/1126-6708/2009/08/090}
  {\path{doi:10.1088/1126-6708/2009/08/090}}.

\bibitem{Dorenbosch:1986tb}
J.~Dorenbosch, et~al., {Experimental verification of the universality of
  electron-neutrino and muon-neutrino coupling to the neutral weak current},
  Phys.Lett. B180 (1986) 303.
\newblock \href {http://dx.doi.org/10.1016/0370-2693(86)90315-1}
  {\path{doi:10.1016/0370-2693(86)90315-1}}.

\bibitem{Allaby:1987vr}
J.~Allaby, et~al., {A Precise Determination of the Electroweak Mixing Angle
  from Semileptonic Neutrino Scattering}, Z.Phys. C36 (1987) 611.
\newblock \href {http://dx.doi.org/10.1007/BF01630598}
  {\path{doi:10.1007/BF01630598}}.

\bibitem{Blondel:1989ev}
A.~Blondel, P.~Bockmann, H.~Burkhardt, F.~Dydak, A.~Grant, et~al., {Electroweak
  parameters from a high statistics neutrino nucleon scattering experiment},
  Z.Phys. C45 (1990) 361--379.
\newblock \href {http://dx.doi.org/10.1007/BF01549665}
  {\path{doi:10.1007/BF01549665}}.

\bibitem{Zeller:2001hh}
G.~Zeller, et~al., {A Precise determination of electroweak parameters in
  neutrino nucleon scattering}, Phys.Rev.Lett. 88 (2002) 091802.
\newblock \href {http://arxiv.org/abs/hep-ex/0110059}
  {\path{arXiv:hep-ex/0110059}}, \href
  {http://dx.doi.org/10.1103/PhysRevLett.88.091802}
  {\path{doi:10.1103/PhysRevLett.88.091802}}.

\bibitem{Gonzalez-Garcia:2014fba}
M.~C. Gonzalez-Garcia, M.~Maltoni, T.~Schwetz, {Updated fit to three neutrino
  mixing: status of leptonic CP violation}, JHEP 11 (2014) 052.
\newblock \href {http://arxiv.org/abs/1409.5439} {\path{arXiv:1409.5439}},
  \href {http://dx.doi.org/10.1007/JHEP11(2014)052}
  {\path{doi:10.1007/JHEP11(2014)052}}.

\bibitem{nufit-2.0}
M.~Gonzalez-Garcia, M.~Maltoni, T.~Schwetz, {NuFit 2.0 (2014)},
  \href{http://www.nu-fit.org}{\tt http://www.nu-fit.org}.

\bibitem{Palladino:2015vna}
A.~Palladino, F.~Vissani, {The natural parameterization of cosmic neutrino
  oscillations}, Eur. Phys. J. C75 (2015) 433.
\newblock \href {http://arxiv.org/abs/1504.05238} {\path{arXiv:1504.05238}},
  \href {http://dx.doi.org/10.1140/epjc/s10052-015-3664-6}
  {\path{doi:10.1140/epjc/s10052-015-3664-6}}.

\end{thebibliography}

\end{document}